\documentstyle[12pt]{article}

\catcode`\@=11
\long\def\@makefntext#1{ 
\protect\noindent \hbox to 3.2pt {\hskip-.9pt
$^{{\ninerm\@thefnmark}}$\hfil}#1\hfill} 

\def\thefootnote{\fnsymbol{footnote}}
 \def\@makefnmark{\hbox to 0pt{$^{\@thefnmark}$\hss}}  

\def\ps@myheadings{\let\@mkboth\@gobbletwo
\def\@oddhead{\hbox{} 
\rightmark\hfil\ninerm\thepage}
\def\@oddfoot{}\def\@evenhead{\ninerm\thepage\hfil 
\leftmark\hbox{}}\def\@evenfoot{}
\def\sectionmark##1{}\def\subsectionmark##1{}}

\begin{document}

\newcommand{\symbolfootnote}{\renewcommand{\thefootnote}
	{\fnsymbol{footnote}}}
\renewcommand{\thefootnote}{\fnsymbol{footnote}}
\newcommand{\alphfootnote}
	{\setcounter{footnote}{0}
	 \renewcommand{\thefootnote}{\sevenrm\alph{footnote}}}

\newcounter{sectionc}\newcounter{subsectionc}\newcounter{subsubsectionc}
\renewcommand{\section}[1] {\vspace{0.6cm}\addtocounter{sectionc}{1}
\setcounter{subsectionc}{0}\setcounter{subsubsectionc}{0}\noindent
	{\bf\thesectionc. #1}\par\vspace{0.4cm}}
\renewcommand{\subsection}[1] {\vspace{0.6cm}\addtocounter{subsectionc}{1}
	\setcounter{subsubsectionc}{0}\noindent
	{\it\thesectionc.\thesubsectionc. #1}\par\vspace{0.4cm}}
\renewcommand{\subsubsection}[1]
{\vspace{0.6cm}\addtocounter{subsubsectionc}{1}
	\noindent {\rm\thesectionc.\thesubsectionc.\thesubsubsectionc.
	#1}\par\vspace{0.4cm}}
\newcommand{\nonumsection}[1] {\vspace{0.6cm}\noindent{\bf #1}
	\par\vspace{0.4cm}}

\newcounter{appendixc}
\newcounter{subappendixc}[appendixc]
\newcounter{subsubappendixc}[subappendixc]
\renewcommand{\thesubappendixc}{\Alph{appendixc}.\arabic{subappendixc}}
\renewcommand{\thesubsubappendixc}
	{\Alph{appendixc}.\arabic{subappendixc}.\arabic{subsubappendixc}}

\renewcommand{\appendix}[1] {\vspace{0.6cm}
        \refstepcounter{appendixc}
        \setcounter{figure}{0}
        \setcounter{table}{0}
        \setcounter{equation}{0}
        \renewcommand{\thefigure}{\Alph{appendixc}.\arabic{figure}}
        \renewcommand{\thetable}{\Alph{appendixc}.\arabic{table}}
        \renewcommand{\theappendixc}{\Alph{appendixc}}
        \renewcommand{\theequation}{\Alph{appendixc}.\arabic{equation}}
        \noindent{\bf Appendix \theappendixc #1}\par\vspace{0.4cm}}
\newcommand{\subappendix}[1] {\vspace{0.6cm}
        \refstepcounter{subappendixc}
        \noindent{\bf Appendix \thesubappendixc. #1}\par\vspace{0.4cm}}
\newcommand{\subsubappendix}[1] {\vspace{0.6cm}
        \refstepcounter{subsubappendixc}
        \noindent{\it Appendix \thesubsubappendixc. #1}
	\par\vspace{0.4cm}}

\def\abstracts#1{{
	\centering{\begin{minipage}{30pc}\tenrm\baselineskip=12pt\noindent
	\centerline{\tenrm ABSTRACT}\vspace{0.3cm}
	\parindent=0pt #1
	\end{minipage} }\par}}

\newcommand{\bibit}{\it}
\newcommand{\bibbf}{\bf}
\renewenvironment{thebibliography}[1]
	{\begin{list}{\arabic{enumi}.}
	{\usecounter{enumi}\setlength{\parsep}{0pt}
\setlength{\leftmargin 1.25cm}{\rightmargin 0pt}
	 \setlength{\itemsep}{0pt} \settowidth
	{\labelwidth}{#1.}\sloppy}}{\end{list}}

\topsep=0in\parsep=0in\itemsep=0in
\parindent=1.5pc

\newcounter{itemlistc}
\newcounter{romanlistc}
\newcounter{alphlistc}
\newcounter{arabiclistc}
\newenvironment{itemlist}
    	{\setcounter{itemlistc}{0}
	 \begin{list}{$\bullet$}
	{\usecounter{itemlistc}
	 \setlength{\parsep}{0pt}
	 \setlength{\itemsep}{0pt}}}{\end{list}}

\newenvironment{romanlist}
	{\setcounter{romanlistc}{0}
	 \begin{list}{$($\roman{romanlistc}$)$}
	{\usecounter{romanlistc}
	 \setlength{\parsep}{0pt}
	 \setlength{\itemsep}{0pt}}}{\end{list}}

\newenvironment{alphlist}
	{\setcounter{alphlistc}{0}
	 \begin{list}{$($\alph{alphlistc}$)$}
	{\usecounter{alphlistc}
	 \setlength{\parsep}{0pt}
	 \setlength{\itemsep}{0pt}}}{\end{list}}

\newenvironment{arabiclist}
	{\setcounter{arabiclistc}{0}
	 \begin{list}{\arabic{arabiclistc}}
	{\usecounter{arabiclistc}
	 \setlength{\parsep}{0pt}
	 \setlength{\itemsep}{0pt}}}{\end{list}}

\newcommand{\fcaption}[1]{
        \refstepcounter{figure}
        \setbox\@tempboxa = \hbox{\tenrm Fig.~\thefigure. #1}
        \ifdim \wd\@tempboxa > 6in
           {\begin{center}
        \parbox{6in}{\tenrm\baselineskip=12pt Fig.~\thefigure. #1 }
            \end{center}}
        \else
             {\begin{center}
             {\tenrm Fig.~\thefigure. #1}
              \end{center}}
        \fi}

\newcommand{\tcaption}[1]{
        \refstepcounter{table}
        \setbox\@tempboxa = \hbox{\tenrm Table~\thetable. #1}
        \ifdim \wd\@tempboxa > 6in
           {\begin{center}
        \parbox{6in}{\tenrm\baselineskip=12pt Table~\thetable. #1 }
            \end{center}}
        \else
             {\begin{center}
             {\tenrm Table~\thetable. #1}
              \end{center}}
        \fi}

\def\@citex[#1]#2{\if@filesw\immediate\write\@auxout
	{\string\citation{#2}}\fi
\def\@citea{}\@cite{\@for\@citeb:=#2\do
	{\@citea\def\@citea{,}\@ifundefined
	{b@\@citeb}{{\bf ?}\@warning
	{Citation `\@citeb' on page \thepage \space undefined}}
	{\csname b@\@citeb\endcsname}}}{#1}}

\newif\if@cghi
\def\cite{\@cghitrue\@ifnextchar [{\@tempswatrue
	\@citex}{\@tempswafalse\@citex[]}}
\def\citelow{\@cghifalse\@ifnextchar [{\@tempswatrue
	\@citex}{\@tempswafalse\@citex[]}}
\def\@cite#1#2{{$\null^{#1}$\if@tempswa\typeout
	{IJCGA warning: optional citation argument
	ignored: `#2'} \fi}}
\newcommand{\citeup}{\cite}

\def\fnm#1{$^{\mbox{\scriptsize #1}}$}
\def\fnt#1#2{\footnotetext{\kern-.3em
	{$^{\mbox{\sevenrm #1}}$}{#2}}}

\font\twelvebf=cmbx10 scaled\magstep 1
\font\twelverm=cmr10 scaled\magstep 1
\font\twelveit=cmti10 scaled\magstep 1
\font\elevenbfit=cmbxti10 scaled\magstephalf
\font\elevenbf=cmbx10 scaled\magstephalf
\font\elevenrm=cmr10 scaled\magstephalf
\font\elevenit=cmti10 scaled\magstephalf
\font\bfit=cmbxti10
\font\tenbf=cmbx10
\font\tenrm=cmr10
\font\tenit=cmti10
\font\ninebf=cmbx9
\font\ninerm=cmr9
\font\nineit=cmti9
\font\eightbf=cmbx8
\font\eightrm=cmr8
\font\eightit=cmti8

\newcommand{\bea}{\begin{eqnarray}}
\newcommand{\eea}{\end{eqnarray}}
\newcommand{\beq}{\begin{equation}}
\newcommand{\eeq}{\end{equation}}
\newcommand{\nnel}{\nonumber \\ {}}

\newcommand{\tselea}[1]{\label{#1}}
\newcommand{\tseleq}[1]{\label{#1}}
\newcommand{\tbib}[1]{\bibitem{#1}}
\newcommand{\tref}[1]{(\ref{#1})}
\newcommand{\tcite}[1]{\cite{#1}}


\newcommand{\auth}[1]{{ #1}}
\newcommand{\tsetit}[1]{{\it #1}}
\newcommand{\journal}[1]{{\it{#1}}}
\newcommand{\vol}[1]{{\bf #1}}
\newcommand{\yr}[1]{(19#1)}
\newcommand{\NP}{\journal{Nucl. Phys.}}
\newcommand{\PR}{\journal{Phys. Rev.}}
\newcommand{\PRL}{\journal{Phys. Rev. Lett.}}
\newcommand{\PL}{\journal{Phys. Lett.}}
\newcommand{\AP}{\journal{Ann. Phys.}}
\newcommand{\JMP}{\journal{J. Math. Phys.}}

\newcommand{\tnote}[1]{}

\newcommand{\tseno}[1]{}

\newcommand{\bra}[1]{\langle  #1  |}
\newcommand{\ket}[1]{| #1 \rangle}
\newcommand{\ebh}{e^{-\beta H} }
\newcommand{\ttr}[1]{{\rm Tr} \{ {#1} \}}
\newcommand{\texpect}[1]{\ttr{ \ebh #1} /\ttr{ \ebh}}
\newcommand{\texp}[1]{\langle \langle #1 \rangle \rangle}
\newcommand{\real}[1]{{\rm Re} \{ #1 \}}
\newcommand{\imag}[1]{{\rm Im} \{ #1 \}}


\begin{flushright}
Imperial/TP/93-94/26 \\
hep-ph/9404262 \\
15th April, 1994 \\
\end{flushright}


\centerline{\tenbf WHAT IS BEING CALCULATED WITH}
\baselineskip=16pt
\centerline{\tenbf THERMAL FIELD THEORY?
\footnote{Talk
given at the Lake Louise Winter Institute on Particle Physics and Cosmology,
20th - 26th February, 1994, Lake Louise, Canada.
}
}
\vspace{0.8cm}
\centerline{\tenrm T.S. EVANS\footnote{E-mail: T.Evans@IC.AC.UK}}
\baselineskip=13pt
\centerline{\tenit Blackett Lab., Imperial College, Prince Consort Road}
\baselineskip=12pt
\centerline{\tenit London, SW7 2BZ, U.K.}
\vspace{0.9cm}
\abstracts{Thermal field theory is reviewed briefly.  It is noted
that until recently it was not known what type of real-time Green function
was being calculated in the Euclidean approach.  The formal answer to
this question is then given and the unexpectedly complicated answer
is discussed.  The physical implications of the results are then considered.
}

\vspace{0.8cm}

\twelverm   
\baselineskip=14pt


\section{Introduction} \label{sintro}
Thermal field theory, or finite temperature field theory as it is
also called, has been around for at least forty years\tcite{Ma}.
The resulting tools, Feynman rules etc., are very similar to those
found in `normal' (zero temperature) quantum field theory, and which has
been used very successfully in particle physics and elsewhere.
However thermal field theory encodes additional physics not present in zero
temperature field theory, and the two are quite different in many ways.


Thus, the key message of this work is that one should not be seduced
into thinking thermal field theory is `just' like ordinary field
theory.  Hence one {\em must} think carefully about the link
between a calculation of Green functions etc. and a given physical
problem and one must not rely upon the superficial similarities
between the zero-temperature  and thermal cases.

\subsection{What is thermal field theory?}

Thermal field theory is the combination of two theories.  The first
element is quantum field theory which is used to describe the
behaviour of small, often fundamental, particles.  One uses it to
describe the dynamics of just of few such particles e.g. as in
e$^+$-e$^-$ collisions at LEP.  A fundamental aspect of the physics
encoded in this theory is the idea of quantum fluctuations, as
characterised by Heisenberg's uncertainty relation.  The effects of
such quantum fluctuations on physical processes is represented by
Feynman diagrams with internal lines such as in figure \ref{fese}.
The second theory is thermodynamics/statistical mechanics, which is
used when one wants to study many body problems.  This involves the
ideas of statistical fluctuations, as seen in the appearance of
Bose-Einstein, $n_b$, or Fermi-Dirac, $n_f$, distributions.

The combination of quantum field theory and statistical mechanics is
called thermal field theory.  It is therefore used in many-body
problems where the underlying dynamics is described by quantum
fields.  It is describing quite a different physical
situation as compared to that appropriate for normal quantum field
theory.  Both quantum and statistical fluctuations must be
accounted for at the same time.  It is therefore not surprising if
thermal field theory has some quite different properties from
standard quantum field theory.  What is amazing is that formally the two can
be described in a very similar way, as I will briefly
outline below.

As an example of the different physics,
consider an electron propagating through a
vacuum and through a QED plasma.
The usual quantum fluctuations are always present
and must be accounted for.  The first Feynman diagram of figure
\ref{fese} is one such contribution and the internal lines represent
interactions of the electron with virtual particles present in the
vacuum.  For the many-body problem only, one must also consider
interactions with the real physical particles present in the plasma.
The last diagram of figure \ref{fese}
represents such an example, while
the middle two diagrams represent the effects of a mixture of quantum
and statistical fluctuations.
\typeout{*** electron self-energy diagrams}
\begin{figure}[htb]
\begin{center}
\setlength{\unitlength}{2pt}%
\begin{picture}(50,11)(-11,-3)
\thicklines
\put(-11,0){\vector(1,0){6}}
\put(-5,0){\line(1,0){5}}
\put(22,0){\vector(1,0){8}}
\put(30,0){\line(1,0){3}}
\put(0,0){\circle*{3}}
\put(22,0){\circle*{3}}
\put(2,0){\oval(4,4)[tl]}
\put(2,4){\oval(4,4)[br]}
\multiput(4,4)(-0.1,0.2){5}{\line(1,0){0.1}}
\multiput(3.5,5)(-0.1,0.1){5}{\line(1,0){0.1}}
\multiput(3,5.5)(-0.1,0.2){5}{\line(1,0){0.1}}
\put(4.5,6.5){\oval(4,4)[tl]}
\multiput(4.5,8.5)(0.2,-0.1){5}{\line(1,0){0.1}}
\multiput(5.5,8)(0.1,-0.1){5}{\line(1,0){0.1}}
\multiput(6,7.5)(0.2,-0.1){5}{\line(1,0){0.1}}
\put(7,9){\oval(4,4)[br]}
\put(11,9){\oval(4,4)[t]}
\put(15,9){\oval(4,4)[bl]}
\multiput(15,7)(0.2,0.1){5}{\line(1,0){0.1}}
\multiput(16,7.5)(0.1,0.1){5}{\line(1,0){0.1}}
\multiput(16.5,8)(0.2,0.1){5}{\line(1,0){0.1}}
\put(17.5,6.5){\oval(4,4)[tr]}
\multiput(19.5,6.5)(-0.1,-0.2){5}{\line(1,0){0.1}}
\multiput(19,5.5)(-0.1,-0.1){5}{\line(1,0){0.1}}
\multiput(18.5,5)(-0.1,-0.2){5}{\line(1,0){0.1}}
\put(20,4){\oval(4,4)[bl]}
\put(20,0){\oval(4,4)[tr]}
\put(11,0){\oval(22,22)[b]}
\put(22,0){\line(0,-1){1}}
\end{picture}
\begin{picture}(50,11)(-11,-3)
\thicklines
\put(-11,0){\vector(1,0){6}}
\put(-5,0){\line(1,0){5}}
\put(22,0){\vector(1,0){8}}
\put(30,0){\line(1,0){3}}
\put(0,0){\circle*{3}}
\put(22,0){\circle*{3}}
\put(2,0){\oval(4,4)[tl]}
\put(2,4){\oval(4,4)[br]}
\multiput(4,4)(-0.1,0.2){5}{\line(1,0){0.1}}
\multiput(3.5,5)(-0.1,0.1){5}{\line(1,0){0.1}}
\multiput(3,5.5)(-0.1,0.2){5}{\line(1,0){0.1}}
\put(4.5,6.5){\oval(4,4)[tl]}
\multiput(4.5,8.5)(0.2,-0.1){5}{\line(1,0){0.1}}
\multiput(5.5,8)(0.1,-0.1){5}{\line(1,0){0.1}}
\multiput(6,7.5)(0.2,-0.1){5}{\line(1,0){0.1}}
\put(7,9){\oval(4,4)[br]}
\put(11,9){\oval(4,4)[t]}
\put(15,9){\oval(4,4)[bl]}
\multiput(15,7)(0.2,0.1){5}{\line(1,0){0.1}}
\multiput(16,7.5)(0.1,0.1){5}{\line(1,0){0.1}}
\multiput(16.5,8)(0.2,0.1){5}{\line(1,0){0.1}}
\put(17.5,6.5){\oval(4,4)[tr]}
\multiput(19.5,6.5)(-0.1,-0.2){5}{\line(1,0){0.1}}
\multiput(19,5.5)(-0.1,-0.1){5}{\line(1,0){0.1}}
\multiput(18.5,5)(-0.1,-0.2){5}{\line(1,0){0.1}}
\put(20,4){\oval(4,4)[bl]}
\put(20,0){\oval(4,4)[tr]}
\put(0,0){\vector(1,-2){5}}
\put(22,0){\line(-1,-2){5}}
\put(8,-16){\makebox(0,0)[lb]{$n_f$}}
\put(22,0){\line(0,-1){1}}
\end{picture}
\begin{picture}(50,11)(-11,-3)
\thicklines
\put(-11,0){\vector(1,0){6}}
\put(-5,0){\line(1,0){5}}
\put(22,0){\vector(1,0){8}}
\put(30,0){\line(1,0){3}}
\put(0,0){\circle*{3}}
\put(22,0){\circle*{3}}
\put(2,0){\oval(4,4)[tl]}
\put(2,4){\oval(4,4)[br]}
\multiput(4,4)(-0.1,0.2){5}{\line(1,0){0.1}}
\multiput(3.5,5)(-0.1,0.1){5}{\line(1,0){0.1}}
\multiput(3,5.5)(-0.1,0.2){5}{\line(1,0){0.1}}
\put(4.5,6.5){\oval(4,4)[tl]}
\multiput(4.5,8.5)(0.2,-0.1){5}{\line(1,0){0.1}}
\multiput(5.5,8)(0.1,-0.1){5}{\line(1,0){0.1}}
\multiput(6,7.5)(0.2,-0.1){5}{\line(1,0){0.1}}
\put(8,10){\makebox(0,0)[lb]{$n_b$}}
\multiput(15,7)(0.2,0.1){5}{\line(1,0){0.1}}
\multiput(16,7.5)(0.1,0.1){5}{\line(1,0){0.1}}
\multiput(16.5,8)(0.2,0.1){5}{\line(1,0){0.1}}
\put(17.5,6.5){\oval(4,4)[tr]}
\multiput(19.5,6.5)(-0.1,-0.2){5}{\line(1,0){0.1}}
\multiput(19,5.5)(-0.1,-0.1){5}{\line(1,0){0.1}}
\multiput(18.5,5)(-0.1,-0.2){5}{\line(1,0){0.1}}
\put(20,4){\oval(4,4)[bl]}
\put(20,0){\oval(4,4)[tr]}
\put(11,0){\oval(22,22)[b]}
\put(22,0){\line(0,-1){1}}
\end{picture}
\begin{picture}(50,11)(-11,-3)
\thicklines
\put(-11,0){\vector(1,0){6}}
\put(-5,0){\line(1,0){5}}
\put(22,0){\vector(1,0){8}}
\put(30,0){\line(1,0){3}}
\put(0,0){\circle*{3}}
\put(22,0){\circle*{3}}
\put(0,0){\vector(1,-2){5}}
\put(22,0){\line(-1,-2){5}}
\put(8,-16){\makebox(0,0)[lb]{$n_f$}}
\put(2,0){\oval(4,4)[tl]}
\put(2,4){\oval(4,4)[br]}
\multiput(4,4)(-0.1,0.2){5}{\line(1,0){0.1}}
\multiput(3.5,5)(-0.1,0.1){5}{\line(1,0){0.1}}
\multiput(3,5.5)(-0.1,0.2){5}{\line(1,0){0.1}}
\put(4.5,6.5){\oval(4,4)[tl]}
\multiput(4.5,8.5)(0.2,-0.1){5}{\line(1,0){0.1}}
\multiput(5.5,8)(0.1,-0.1){5}{\line(1,0){0.1}}
\multiput(6,7.5)(0.2,-0.1){5}{\line(1,0){0.1}}
\put(8,10){\makebox(0,0)[lb]{$n_b$}}
\multiput(15,7)(0.2,0.1){5}{\line(1,0){0.1}}
\multiput(16,7.5)(0.1,0.1){5}{\line(1,0){0.1}}
\multiput(16.5,8)(0.2,0.1){5}{\line(1,0){0.1}}
\put(17.5,6.5){\oval(4,4)[tr]}
\multiput(19.5,6.5)(-0.1,-0.2){5}{\line(1,0){0.1}}
\multiput(19,5.5)(-0.1,-0.1){5}{\line(1,0){0.1}}
\multiput(18.5,5)(-0.1,-0.2){5}{\line(1,0){0.1}}
\put(20,4){\oval(4,4)[bl]}
\put(20,0){\oval(4,4)[tr]}
\end{picture}
\end{center}
\vspace{1cm}
\fcaption{Different types of process which occur when in a heat bath.}
\label{fese}
\end{figure}
The graphical representation of figure \ref{fese} is closely linked
to terms seen when the Feynman integral representation of the
electron self-energy for Minkowskii time-ordered Green functions
in thermal field theory is examined\cite{We}.

\subsection{An example of the problem}\label{sexprob}

Consider the example of a decay rate of a
particle in a many-body situation, say a
baryon number violating process in the early universe.
Typically in a one-loop Feynman diagram calculation, the decay rate
at high temperatures is found to be of order $g^2 T$ where $g$ is
the gauge coupling and $T$ the temperature.

However we know from our experience with zero-temperature field
theory that the coupling we measure at lab scales is not always a
good measure of the coupling strength relevant for processes at
other scales.  One excellent solution is to use renormalisation
group which gives a link between the best values to use for the coupling
constant at different scales.
It is therefore vital to supplement the original
perturbative calculation with what is non-perturbative information
obtained from renormalisation group analysis.

In a many-body problem, it is therefore essential to
use the renormalisation group to `run' the coupling from the known
lab (zero temperature) value,
to a value
appropriate for a hot plasma, effectively $g \rightarrow g(T)$.
This can be done in
principle and one needs to know how the quantum and statistical
corrections to the coupling.  This usually involves calculating
one-loop three-point diagrams.

Now the problems begin to appear.  The
calculations have been done for QCD but the results are generic.
There are two distinct calculational schemes used
in thermal field theory, RTF (the Real-Time Formalism) and ITF (the
Imaginary Time Formalism).  In this
case the calculations using RTF suggested that such diagrams could
grow as fast as $T^3$ at high temperatures\tcite{RTF3QCD}
for the case where the external legs of the diagram
were at zero energy.  On the other  hand, it was known that if one
used ITF to do the calculation,  with zero energy on external legs,
the diagrams\tcite{FT} can not grow faster than $T$.
In solving this apparent disparity, a proper understanding of the
ITF calculations in particular emerges.


\section{A general approach to thermal field theory}

In order to provide a precise answer to the question of what is
being calculated in the different thermal field theory schemes, I
will first outline what these schemes are.  I will use a path
ordered approach to thermal field theory\tcite{Raybook,Ka,LvW}  as
opposed to the alternative operator style used in Thermo Field
Dynamics\tcite{TFD,LvW}.  The results hold in either case.  The
great advantage of this path ordered approach is that we can obtain
different formalisms through different choices of a curve $C$ in the
complex time plane.   The derivation will show that they must
contain the same physics, they merely encode it in different ways.

The starting point is the generating functional $Z[j]$
\beq
Z[j] :=
\ttr{\ebh T_c\exp [ i\int_Cd\tau\int d^3\vec{x}
\; j(\tau,\vec{x})\phi(\tau,\vec{x}) ] }
\tseleq{eZjdef}
\eeq
for the example of a single scalar field.
The $T_c$ indicates some sort of time ordering
of the operators, and the time integration is along some
path $C$, both to be discussed later.
Note that the statistical mechanics element is suggesting
the $\ttr{ \ebh \ldots }$,
so that $Z[j=0]$ is the partition function of statistical
mechanics.  The quantum field theory comes in through the
path ordered exponential.  This means that the Green functions, the key
tools of Quantum field theory, are generated by
taking functional derivatives with respect to the sources of the generating
functional and then setting these unphysical sources to zero
\bea
\left.  \frac{\partial^N Z[j] }
{\partial j_1(\tau_1)\partial j_2(\tau_2)\ldots \partial j_N(\tau_N)}
\right|_{j=0} &=&
\ttr { \ebh
T_c \phi_1(\tau_1)\phi _2(\tau_2)...\phi_N(\tau_N) }
\tselea{egfgenc}
\eea

Now the problem in trying to link the TFT expression \tref{eZjdef}
to those normally encountered in regular QFT is the thermal
weight $\ebh$.  The way this is formally accounted for is the
{\bf\sc key} trick of thermal field theory, and that is
to regard the thermal weight as a time evolution operator.  We
have
\bea
\ttr{ \ebh \ldots }&=&
\sum_{n} \bra{n;\tau_{in}} \ebh \ldots \ket{n;\tau_{in}} =
\sum_{n} \bra{n;\tau_{in}-i \beta} \ldots \ket{n;\tau_{in}}    .
\tselea{ekey}
\eea
Thus we are treating
this thermal weight as if it was a time translation of amount $\beta$ in
the Euclidean direction.  Note that temperature has not replaced
time in any sense, the dots in \tref{ekey} represent combinations of field
operators which still have any time lying on the path $C$.

This can now be compared with the expressions usually encountered in
quantum field theory which are of the form
$\bra{n_{out};\tau_{out}} \ldots \ket{n_{in};\tau_{in}}$.  One
can then just use the standard tricks of quantum field theory
to proceed.  For instance, the path integral can be used and this
gives\tnote{\tcite{Raybook},
(4.102) except it is for $j=0$, but see\tcite{LvW}, pp156 for
$\texpect{F[\phi]}$.}
\bea
Z[j] &=&
\sum_{\Phi_{in}} \bra{ \Phi_{in}(\vec{x});\tau_{in} - i \beta}
\exp\{T_C \int_C d\tau \int d^3\vec{x} \;
j(\tau,\vec{x}) \hat{\phi}(\tau,\vec{x}) \}
\ket{ \Phi_{in}(\vec{x});\tau_{in} }
\nnel
&=&
\int_{PBC}  D\Phi \; \exp \{i \int_C d\tau \int_V d^3\vec{x}
( {\cal L} + j \Phi ) \}
\tselea{eZjpi}
\eea
where $C$ must now run from $\tau_{in}$ to $\tau_{out}=\tau_{in}-i\beta$.
The $PBC$ on the path integral stands for the periodic boundary
conditions which must be enforced, namely that
$\Phi(\vec{x},\tau_{in} )=\Phi(\vec{x},\tau_{in}-i\beta
)=\Phi_{in}(\vec{x})$.

Equation \tref{eZjpi} is a remarkably compact
result.  It is also clear that temperature only appears as a
boundary condition, viz we only consider fields configurations which
are periodic in time over $-i\beta$.  This does not stop us from
having field configurations with any sort of time dependence in the
real physical time direction or over the unphysical range of
imaginary times up to $-i\beta$.

For the purposes of the problem at hand, the most important point is
that all times must lie on a curve $C$ which runs from $\tau_{in}$ to
$\tau_{out}=\tau_{in}-i \beta$.  The fields in this formalism, e.g.
the operators in Green functions generated from $Z[j]$, only take
times whose values lie on $C$.  Further, working from the definition
of the path integral, we see that the path ordering in \tref{eZjdef}
{\em must} be path ordering with respect to this same curve $C$.
Thus the Green functions generated from $Z[j]$ \tref{egfgenc} are
ones where the field operators are ordered according to the position
of their time arguments on $C$, the closer to $\tau_{in}$
($\tau_{out}$) the closer that field is placed to the left (right) in
the thermal expectation value.  Thus we have identified for the
general case exactly what is being calculated in thermal field
theories.

The great advantage of this path ordered approach to thermal field
theory is that we are
free to choose $C$ as we wish so long as it runs from   $\tau_{in}$
to $\tau_{out}=\tau_{in}-i \beta$.  Thus we can obtain the two main
 formalisms through different choices of $C$.  The
derivation shows that all formalisms must contain the same physics, they
merely encode it in different ways as they generate different types
of Green function, path ordered with respect to different curves.

\section{The Real-Time Formalism}

The RTF (real-time formalism) can be generated by choosing the
curve\tcite{TSEnrtf} of figure \ref{fnrtf},
\typeout{LaTeX figure FNRTF}
\begin{figure}[htb]
\setlength{\unitlength}{0.4pt}
\begin{picture}(500,280)(100,315)
\thicklines
\put(160,340){\circle*{10}}
\put(160,530){\circle*{10}}
\put(540,530){\circle*{10}}
\put(320,320){\vector( 0, 1){270}}
\put(120,525){\vector( 1, 0){450}}
\put(160,530){\vector( 1, 0){240}}
\put(400,530){\line( 1, 0){140}}
\put(540,530){\vector(-2,-1){180}}
\put(360,440){\vector(-2,-1){200}}
\put(320,420){\circle{10}}
\put(170,435){\makebox(0,0)[lb]{$-i(1-\alpha)\beta$}}
\put(135,315){\makebox(0,0)[lb]{$-t_{in}- i \beta$}}
\put(135,545){\makebox(0,0)[lb]{$-t_{in}$}}
\put(480,545){\makebox(0,0)[lb]{$(1-1/\alpha)t_{in}$}}
\put(305,500){\makebox(0,0)[lb]{$0$}}
\put(525,495){\makebox(0,0)[lb]{$\Re e (\tau)$}}
\put(230,560){\makebox(0,0)[lb]{$\Im m (\tau)$}}
\put(390,540){\makebox(0,0)[lb]{\large $C_{1}$}}
\put(370,423){\makebox(0,0)[lb]{\large $C_{2}$}}
\end{picture}
\typeout{LaTeX figure FNRTF for infinity}
\begin{picture}(400,280)(100,315)
\thicklines
\put(320,320){\vector( 0, 1){270}}
\put(120,525){\vector( 1, 0){450}}
\put(160,530){\vector( 1, 0){240}}
\put(400,530){\line( 1, 0){140}}
\put(540,420){\vector(-1, 0){140}}
\put(400,420){\line(-1, 0){240}}
\put(320,420){\circle{10}}
\put(305,500){\makebox(0,0)[lb]{$0$}}
\put(170,435){\makebox(0,0)[lb]{$-i(1-\alpha)\beta$}}
\put(525,495){\makebox(0,0)[lb]{$\Re e (\tau)$}}
\put(230,560){\makebox(0,0)[lb]{$\Im m (\tau)$}}
\put(390,540){\makebox(0,0)[lb]{\large $C_{1}$}}
\put(370,433){\makebox(0,0)[lb]{\large $C_{2}$}}
\end{picture}
\fcaption{Curves used in RTF.}
\label{fnrtf}
\end{figure}
and then taking the limit of
$t_{in} \rightarrow \infty$ to give the second version in
figure \ref{fnrtf}.  The curve of figure \ref{fnrtf} curve avoids
various problems found\tcite{TSEACPban} when alternative curves
\tcite{NS,LvW} are used.  The parameter $\alpha$ is unphysical and
does not appear in any physical quantity.

The RTF is designed to be very like standard Minkowskii zero
temperature field theory.  When a field lies on the $C_1$ section it
is just like those encountered in Minkowskii
field theory.  Even when the field lies on
$C_2$, the field still depends only on a real-time parameter which
varies between $-\infty$ and $+\infty$.  The formalism contains
fields at real physical times at the expense of having to give
every physical field (the bits associated with $C_1$ time arguments,
`type 1')  a thermal partner (associated with the $C_2$ section,
`type 2').  With experience, it is easy to cope with this doubling.

The important element for the problem at hand is that this means
that if we look at diagrams with type 1 external legs this
corresponds to calculating
\beq
\ttr {\ebh
T_c \phi_1(\tau_1)\phi_2(\tau_2)...\phi_N(\tau_N) }
=\ttr{ \ebh
T \phi_1(t_1)\phi_2(t_2)...\phi_N(t_N) }
 ,
\eeq
where $ \tau_i=t_i \in \Re e$.
RTF was intended to look like zero temperature field
theory, so it is not surprising to see that it is calculating
time-ordered Green functions of fields with real physical times.
This is what is calculated in most RTF calculations.

\section{The Imaginary-Time Formalism}

It turns out that the older 
ITF\tcite{Ma,Raybook,Ka,LvW} (Imaginary-Time
Formalism) is the one
which had not been studied in depth.  The path chosen for ITF in path
ordered approach just runs down the
imaginary-time axis as shown in figure \ref{fitf}.
\typeout{figure: Standard ITF curve}
\begin{figure}[htb]
\begin{center}
\setlength{\unitlength}{.2in}
\begin{picture}(5,5.5)
\thicklines
\put(-0.5,4.6){\vector(1,0){6.1}}
\put(4.8,3.8){$\Re e \, (\tau) $ }
\put(2.5,0.0){\vector(0,1){5.0}}
\put(2.6,5.0){$\Im m \, (\tau) $ }
\thinlines
\put(2.3,4.5){\circle*{0.2}}
\put(2.3,4.5){\vector(0,-1){2.0}}
\put(2.3,2.5){\line(0,-1){2.0}}
\put(2.3,.5){\circle*{0.2}}
\put(1.5,2.5){\large $C$}
\put(2,5){\makebox(0,0){$0$}}
\put(1.5,.5){\makebox(0,0){$-\imath \beta$}}
\end{picture}
\end{center}
\fcaption{Path used for the Imaginary Time Formalism.}
\label{fitf}
\end{figure}
This means that
only Euclidean times appear directly in this formalism.  This is fine
for time independent quantities such as specific heats and other
thermodynamic things.

However, when one wants to learn about real dynamics we need physical,
that is real, times.
ITF, however, gives N-point Green functions which are
functions of $N-1$ independent Euclidean times.  Just like any other
function one may attempt to analytically continue it to real values.
This is usually straightforward given an explicit expression.  One
question that immediately comes to mind is what type of Green
function does one obtain after such a continuation?  After all one
starts with a Green function which is the expectation value of fields
ordered according to their Euclidean time arguments.  In what order
are the fields when all their values have been continued to real
times?
The answer is surprising and not very obvious.

\subsection{The analytic continuation of Euclidean Green functions}

Such a fundamental question for ITF has been studied for
the most important case, that of the two-point function.  This was
done by Baym and Mermin in 1961\tcite{BM,LvW}.  There they showed that
one obtains the retarded/advanced two-point Green functions.  Further
it is easy to see that the real part of these Green functions, which
is related to the effective thermal mass, is the same as the real part of
the time-ordered Green function.  The two different types of
Green functions do differ by $O(T)$ but only in their imaginary part.

What is surprising is that this work had not been extended.  The
example of the running coupling constant, given in section 1.2, 
shows that understanding higher Green functions is important.  It was
as if it was assumed that the same results (real parts the same,
imaginary parts slightly different) held for higher Green functions.


It is not trivial to generalise the work of Baym and Mermin to N-point
functions.
Further, the results for two-point thermal Green
functions do not generalise in a very obvious way either.
 However their basic approach still works.  One uses three ideas.
First the definitions of the various types of Green function are
written down for a general case, the Euclidean time-ordered Green
functions in particular.
 Next the boundary conditions are encoded in terms of Green functions
with fields in a fixed order, the thermal Wightman
functions.  This is a generalisation of the
KMS\tcite{KMS,LvW} (Kubo-Martin-Schwinger) condition from two- to
N-point functions.   One simply uses the definition of
a thermal Green function as trace, uses the cyclicity and then
interprets the $\exp\{-\beta H\}$ factor as a time translation in the
usual way.  Lastly, one finds that one must
make some assumption about the behaviour of the Green functions at
large real times.

One then obtains a spectral representation of the Green functions that
are calculated in ITF.  This is\tcite{TSEnpt}
\bea
\lefteqn{\Gamma^{(N)}( \{ z \} )
= (\frac{-1}{2\pi})^{N-1} \int_{-\infty}^{\infty}
dk_1\ldots dk_N  .
\delta(k_1 + \ldots + k_N)
\; . \; \left[ \rho_{123 \ldots N} ( \{ k \} ; \beta, \ldots)
\right. }
\nnel
&&\times
\frac{i}{(z_2+z_3+ \ldots +z_N) - (k_2 +k_3 + \ldots + k_N)}
\times \frac{i}{(z_3+ \ldots +z_N) - (k_3 + \ldots + k_N) }
\nnel
 && \; \; \; \; \times \; \; \ldots \; \;
\times \left. \frac{i}{z_N - k_N}
\right] \; \; + \; \; (\mbox{ALL permutations of } (123\ldots N) )
\tselea{eNsp}
\eea
The dependence on the three-momentum or space variables has not been
written explicitly.  The $k$'s are real energy variables and the $z$'s
are complex energy variables.  One of the $k$'s is clearly redundant,
and an $N$-th $z$ complex variable has likewise been introduced
through the complex equation $\sum z_j =0$.  This has been done to
make the expressions symmetric and easy to write down.

All the individual details of the
theory (masses, coupling constants etc.)
and the approximation used to calculate the Green function are
contained in the spectral densities, $\rho$.  The dots in the
arguments represent dependence on these factors.  The spectral
densities are the difference of two thermal Wightman functions,
\bea
\lefteqn{\rho_{123 \ldots N}(\{k\}; \beta, \ldots) =
\texpect{ \phi_1(k_1) \phi_2(k_2) \phi_3(k_3) \ldots \phi_N(k_N)} }
\nnel
&& - (-1)^{N}
\texpect{ \phi_N(k_N) \ldots \phi_3(k_3) \phi_2(k_2) \phi_1(k_1) }
\eea
for the case of pure bosonic fields.
For instance in the case of two-point functions one might finds there
is just one spectral density and in simple cases this has the form
$(\theta(k) - \theta(-k) ) \delta(k^2-\omega^2)$.
The $\omega$ is a three-momentum dependent dispersion relation,
$\surd (\vec{k}^2+m^2)$ for relativistic free fields.  In general
the dispersion relation and the spectral function's form are more
complicated\tcite{La,BPY}.

The aim is to extract real dynamics and therefore to look at the
real energy limit of the $z$'s.  However, it is clear that the
expression \tref{eNsp} has cuts on the real $z$ axes.
Further, there are also cuts at real values of $z_1+z_2$,
$z_1+z_2+z_3$, and indeed in all possible sums of these complex
energies.  One must therefore specify from which side of the
cut one is approaching by leaving in a small imaginary
part,
\beq
z_j \rightarrow E_j + i \epsilon_j
\tseleq{eac}
\eeq
where $E_j$ is real, $\epsilon_j$ is real and infinitesimal, and both can
be either positive or negative.  Note that it is the sign of the
$\epsilon_j$'s, and the sign of all possible sums of the
$\epsilon_j$'s, which tell us which side of each cut we are looking
at.  For $N \geq 4$, the sign of the single $\epsilon_j$'s does not
always specify the signs of all possible $\epsilon_j$ sums\tcite{TSEnpt}.

Now we are at last in a position to answer the question exactly what
sort of real-time Green function has ITF calculated, in particular
what operator ordering has we been left with after analytic
continuation.   Every distinct way of approaching all these cuts
corresponds to a different operator ordering and so a different type
of Green function.  To see exactly what type of Green function has
emerged,  one merely does a Fourier transform of  \tref{eNsp} from
$E$'s to real times $t$, for a given analytic continuation \tref{eac}.
However,
the resulting expression can be further manipulated by using
identities such as $\theta(t)+\theta(-t) =1$.  This makes it highly
non trivial to see if an expression is equivalent to something
familiar or not.

The results\tcite{TSEnpt} can be summarised as follows:-
\begin{itemlist}
\item The
analytic continuation of ITF Green functions produces
the thermal {\em Generalised Retarded Functions}.  Eq.s \tref{eNsp} and
\tref{eac}  form a definition of these functions c.f. at zero
temperature\tcite{Ar,ztNpt}.
\item For the special case where one epsilon is positive, say
$\epsilon_j>0$, and all the
others are negative (which fixes all possible sums of epsilons too),
one obtains the $j$-th retarded function, $R_j(t_j; \{ t_{others} \} )$
\tcite{TSEnpt,TSE3pt,BN}.
The advanced functions are obtained when the signs of the epsilons are
the opposite way round.  These functions are discussed
below.
\item {\em No} one analytic continuation of an ITF N-point Green
function produces the time-ordered Green function\tcite{TSEnpt,TSE3pt}.
\end{itemlist}

The retarded and advanced functions are very useful.  They appear in
the linear response approximation which trys to describe small
perturbations to a plasma\tcite{Ka,LvW}.
The retarded functions are usually written as
expectation values of multiple
commutators, with theta functions.  One field is picked out as being
special, they are symmetric only in the remaining fields.  That one special
field must have the latest time  if the retarded function is to be
non-zero.  This means there are $N$ N-point retarded functions.  The
advanced functions are the same but with all the opposite theta
functions $\theta(t) \rightarrow \theta(-t)$ and a possible overall
minus sign.  For instance, for pure bosonic fields,
\bea
R_1(t_1;t_2,t_3,t_4) &=&
\theta(t_1-t_2)\theta(t_2-t_3)\theta(t_3-t_4)
\texpect{ [[[\phi_1,\phi_2],\phi_3],\phi_4]}
\nnel
&& \; \; \; \; + \; \left(\mbox{All (234) permutations} \right)
\eea
More recently, Baier and Niegawa have derived the
link between ITF and the retarded/advanced
functions only, in a much simpler way\tcite{BN}.

One interesting sideline is that I do not know how many of these
generalised functions ITF produces.  There are $N$ retarded and $N$
advanced functions, but they are a small subset of the total number of
generalised retarded functions.  The results to date are shown in table
\tref{tnf}.
\typeout{LLWI - table nf}
\begin{table}[htbp] 
\centering 
\begin{tabular}{c||c|c|c|c|c|c|c|c|c|c}
  $N$          & 2 & 3 & 4  & 5   & 6      & 7         & 8      & 9
&\ldots & $N$
\\ \hline \hline
$R$, $A$       & 2 & 6 & 8  & 10  & 12     & 14        & 16     & 18
& \ldots & $2N$
\\ \hline
RTF            & 4 & 8 & 16 & 32  & 64     & 128       & 256    &
512 & \ldots &
$2^N$ \\ \hline
TWF            & 2 & 6 & 24 & 120 & 720    & 5,040     & 40,320 &
362,880 & \ldots & $N!$
\\ \hline
ITF$\equiv$GRF & 2 & 6 & 32 & 370 & 11,292 & 1,066,044 & 347,326,352
& ? & \ldots &
$>O(N!)$?
\end{tabular}
\tcaption{Numbers of $N$-point functions.
The rows are:- $N$, number of external legs; $R,A$ numbers of
retarded and advanced functions; RTF, numbers of distinct RTF
functions; TWF, the number of thermal Wightman functions;
ITF$\equiv$GRF, the number of generalised retarded functions.
}
\label{tnf}
\end{table}
The results for N=2-6 were produced elsewhere\tcite{TSEnpt}
whereas the $N=7,8$ results were found first by van
Eijck using a different method.  I have cross checked the $N=7$ result.
The results up to $N=6$ were independently derived in the
zero temperature case by Araki\tcite{Ar} but the $N=6$ result quoted
there seems to be incorrect.

\section{Physical Implications}

The most important physical implication comes when we try to compare
time-ordered functions and retarded/advanced functions.  This is
because the former are most easily produced in RTF whereas ITF is
almost always calculating the latter.
One finds for truncated/1PI functions\tcite{TSE3pt}
\beq
\Gamma_{t.o.}^{1PI}(E_1,E_2,E_3) = n_2 n_3 (\sigma_1 R_1^{1PI} -
e^{-\beta E_1} A_1^{1PI} )
+ \left( \mbox{All (123) permutations} \right)
\eeq
where $n_j= (\exp \{\beta E_j\} -\sigma_j)^{-1}$
and $\sigma_j$ is $+1$ ($-1$) if the $j$-th field is bosonic
(fermionic).  $A_j$ is the $j$-th advanced thermal Green function.
For the case of purely bosonic fields, we find that the real parts
of the Green functions in the limit of $T \gg E_j$ satisfy
\bea
\lefteqn{ Re \{\Gamma_{t.o.}^{1PI}(E_1,E_2,E_3) \} }
\nnel
&=&  T^2
\left( Re \{R_1^{1PI}\} \frac{E_1}{E_1 E_2 E_3}
+ \mbox{(All other (123) permutations)} \right) + O(1)
\eea
Roughly speaking then, we find that the time-ordered
($\Gamma_{t.o.}^{1PI}$) and
retarded/advanced 1PI functions differ by a factor of the square of the
temperature.  This accounts for the differences in the results of the
calculations of QCD three-point diagrams\tcite{RTF3QCD,FT}
described in section 1.2. 

{}From this work we see that
the same Feynman diagram represents different types of thermal Green
function for different formalisms.
Further, different types of thermal Green function differ by large amounts.
Therefore it is essential that one understands quite carefully which
type of Green function is required for a given physical problem.
Only then, using the results described here, can one choose the
calculational scheme best suited for your purpose.  The  similarity
in the Feynman rules of thermal and normal zero temperature field
theory obscures the fundamental differences between the physics of
the two theories.  These differences arise because
thermal field theory is modelling a very
different physical situation, a quantum plasma not a quantum
vacuum.

I would like to thank the Royal Society for their support through a
University Research Fellowship.  I have benefited greatly from
discussions with M. van Eijck, F. Gu\'{e}rin and R. Kobes.  Due to
space limitations, I have been unable to do justice to the many
papers in this field.  They can be found by looking at the more
recent and detailed papers listed below.  I would finally like to thank the
organisers for such an enjoyable institute.

\typeout{--- references ---}

\end{document}